# Online Voting with Point to MultiPoint Quantum Key Distribution

Bernardo A. Huberman* and Jing Wang**

*RISE - Research Institutes of Sweden, Stockholm, Sweden

** CableLabs, 858 Coal Creek Circle, Louisville, CO, 80027

## Abstract

The use of quantm mechanisms in the service of voting security suffers from the problem that in order to generate keys for voters and verifiers a point to point connection has to be physically established for each pair, rendering this impractical. We thus propose using Point-to-Multipoint quantum key distribution (QKD) via time division multiplexing (TDM) and wavelength division multiplexing (WDM) in passive optical networks (PON) to improve both the deployment and security of online voting systems.

## Introduction

There are a number of proposals for secure online voting systems that offer a number of required properties, like completeness, privacy and fairness. Typically, these cryptographic voting schemes can be divided into three categories, based on the technique used to anonymize votes.

In schemes based on *homomorphic encryption*, voters submit encrypted votes that are never decrypted. Rather, the submitted ciphertexts are combined to produce a single ciphertext containing the election tally, which is then decrypted.

*Blind signature* schemes split the election authority into an authenticator and tallier. The voter authenticates to the authenticator, presents a blinded vote, and obtains the authenticator's signature on the blinded vote. The voter unblinds the signed vote and submits it via an anonymized channel to the tallier.

And finally, in *mix network* schemes voters authenticate and submit encrypted votes. Votes are anonymized using a mix, and anonymized votes are then decrypted.

All these schemes rely on the use of public and private keys that ensure completeness, privacy and fairness. But recent advances in quantum computing threaten the security of public key encryption which lies at the heart of all these systems.

This is where QKD offers in principle an advantage since the private keys that are generated via quantum mechanisms are provably secure. There are also quantum schemes that rely on entanglement to secure both the vote and the identity of the voters, but the distribution of entangled states among many parties makes them more complicated to implement.

The use of QKD in the service of voting suffers however from a almost fatal problem, ie that in order to generate keys for voters and verifiers a point to point connection has to be physically established for each pair, rendering this impractical.

A solution of this problem is provided by deploying a Point-to-Multipoint quantum key distribution (QKD) via time division multiplexing (TDM) and wavelength division multiplexing (WDM) in passive optical networks (PON) .

This would allow the voting authority to distribute private keys to all the voting participants, who could then use whatever electronic voting they prefer. In particular it would allow Blind signature schemes like those of Fujioka, Okamoto and Ohta, that use bit-commitment and *private keys*.

One way bit-commitment works using private keys is by having two parties, Alice and Bob, perform the following actions:

1. Bob generates a random-bit string, R, and sends it to Alice.

2. Alice creates a message consisting of the bit (i.e the vote) she wishes to commit to, b (it can actually consist of many bits) and of Bob's random string, R. She encrypts it with some random key, K, and sends the result back to Bob. Esubk [R,b].

   That is the commitment phase of the protocol. Bob cannot decrypt the message, so he doesn't know what the bit is. So now,

3. Alice sends Bob the key, K

4. Bob decrypts the message to reveal the bit. He also checks his random key to verify the bit's validity.

For the other voting schemes, private keys would replace the use of the public encryption they are based on, since public keys are being threatened by the advent of quantum computers.

## Proposed Method and Operation Principles

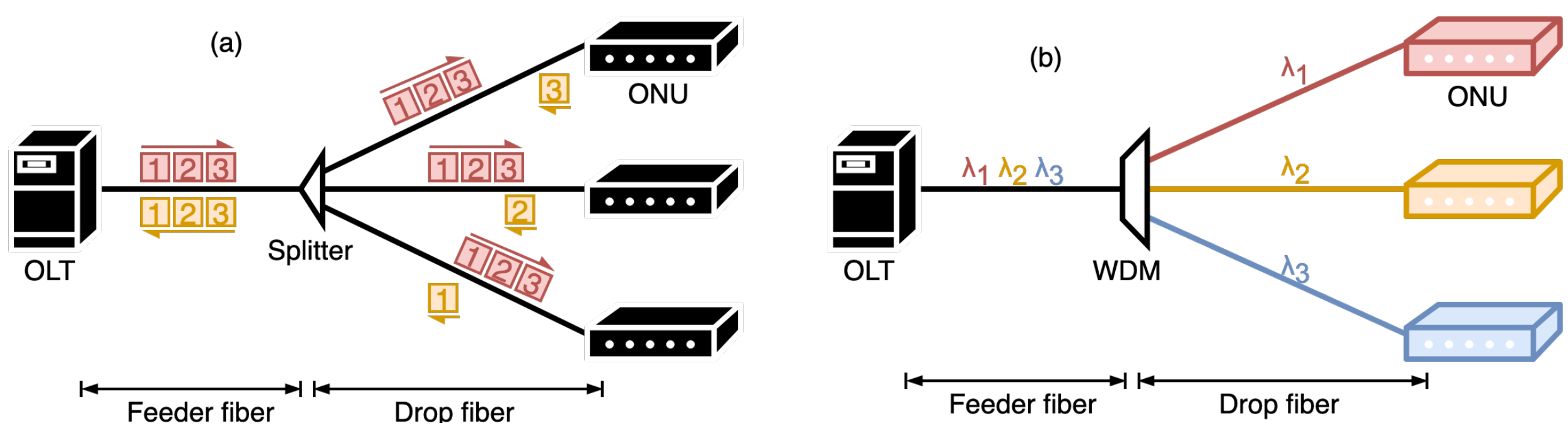

Fig.2 (a) TDM-PON. (b) WDM-PON.

There are two categories of PON architectures, depending on the different methods of user access control. Fig. 2(a) shows the architecture of time-division multiplexing (TDM)-PON, where a power splitter is inserted between the feeder fiber from the optical line terminal (OLT) and drop fibers to optical network units (ONUs). Downstream (DS) data from OLT is broadcasted to all ONUs, and different users rely on data encryption to protect their privacy. The upstream (US) data from ONUs are controlled by time-division multiple access (TDMA), where each ONU is assigned a time slot and packets from different ONUs are multiplexed in the time domain. TDM-PON has been widely deployed due to its low cost and has been standardized in EPON, GPON, 10G-EPON, and XG-PON. Fig. 2(b) shows the architecture of wavelength-division multiplexing (WDM)-PON, where each ONU has a dedicated wavelength. There is no sharing mechanism in either DS or US, and the user access control is much simpler than TDM-PON.

Integrating QKD into PON was first proposed by Townsend et al. in 1994 [6] with downstream deployment, where a single photon source in OLT is shared by multiple ONUs, each equipped with a single photon detector. At the power splitter, since single photons cannot be divided, at most one ONU will receive the photon or the photon cannot reach any user due to the attenuation of splitter. Studies about the coexistence of quantum and classical channels started from Ref [7], which only considered a WDM case in a P2P topology. The integration of QKD in WDM-PON and TDM-PON were demonstrated in [8-9] and [10], respectively. It should be noted that all reported works either only considered either DS or US transmission, or only focused on a P2P topology or limited number of users. In this patent, we proposed a systematic approach to integrated quantum channels into the existing TDM/WDM-PON, with both DS, US, and synchronization channels considered.

The main challenge to integrate QKD into PON is the spontaneous Raman scattering noise that arises when multiplexing quantum channels with classical channels. As an inelastic scattering effect between the incident photons and the medium fiber, Raman scattering changes not only the direction but also the energy of scattered photons. In fiber, the scattered photons can propagate in both directions, and are defined as forward scattering and backscattering, depending on their propagation directions with respect to the incident light. Backscattering is stronger than forward scattering due to its higher efficiency. Photons that lose energy in the scattering are called Stokes photos, which have reduced frequency and red-shifted wavelengths.

Photons that gain energy are called anti-Stokes photons, which have increased frequency and blue-shifted wavelengths. Raman noise cover a spectral range up to 200 nm centered at the wavelength of incident light, with a peak intensity at a frequency shift of 13 THz. Since the scattered photons change their wavelengths, they become noise to existing signals at those wavelengths. Raman scattering noise from classical channels at wavelengths of quantum channels is the dominant factor limiting the QKD distance and key rate.

There are two wavelength choices for QKD; C-band for the lowest fiber loss (0.2 dB/km) or O-band for reduced Raman noise. If quantum channels are assigned to the C-band, it is possible to exploit the standard C-band devices and multiplexing schemes to reduce the system cost. On the other hand, 1310 nm has slightly higher loss (0.33 dB/km) but significantly lower noise, because it is out of the Raman spectrum of most classical channels in the C-band. Moreover, by assigning shorter wavelength to the quantum channel than the classical channels, we can leverage the advantage of weaker anti-Stokes scattering than Stokes scattering. One drawback of O-band is that there is no ITU-T DWDM grid defined yet, and the device cost will be higher. In this work, we will assign quantum channels to around 1310 nm. For all designs in this work, their wavelength plans can be changed without modification to architectures.

In a QKD transmitter (Tx), weak coherent pulses are generated from an attenuated laser to emulate a single photon source. On the receiver (Rx) side, single photon detectors (SPDs) are used to detect single photons. Since SPDs are the most expensive devices in a QKD system, to reduce the overall system cost, we use upstream QKD transmission, where the quantum Tx is located in each ONU and quantum Rx with shared SPDs in the OLT. Since each ONU has its own Tx, it allows each user to exchange individual keys with the OLT at an adjustable rate depending on its channel condition.

**Integration of QKD in TDM-PON**

Fig. 3 shows the architecture and wavelength plan of a TDM-PON with integrated QKD and synchronization (SYNC) channels. The quantum Tx is located in each ONU and a shared quantum Rx in OLT. The quantum channel is at 1310 nm. Classical downstream (DS) and upstream (US) channels are at 1490 nm and 1550 nm respectively, far away from the quantum channel to reduce Raman scattering noise. Since SPDs work in a gated-mode, a low-power low data rate (~100 Mb/s) SYNC channel at 1610 nm is used to provide the trigger signal. It also serves as the auxiliary channel for post-processing, including basis reconciliation, key sifting, error correction, and privacy amplification. In each ONU and OLT, a coarse wavelength-division multiplexer (CWDM) is used to separate/combine classical and quantum channels. Like classical US transmission in TDM-PON, the weak coherent pulses from quantum Tx in ONUs are time interleaved. At any given time, the quantum Rx in OLT only receives photons from one ONU.

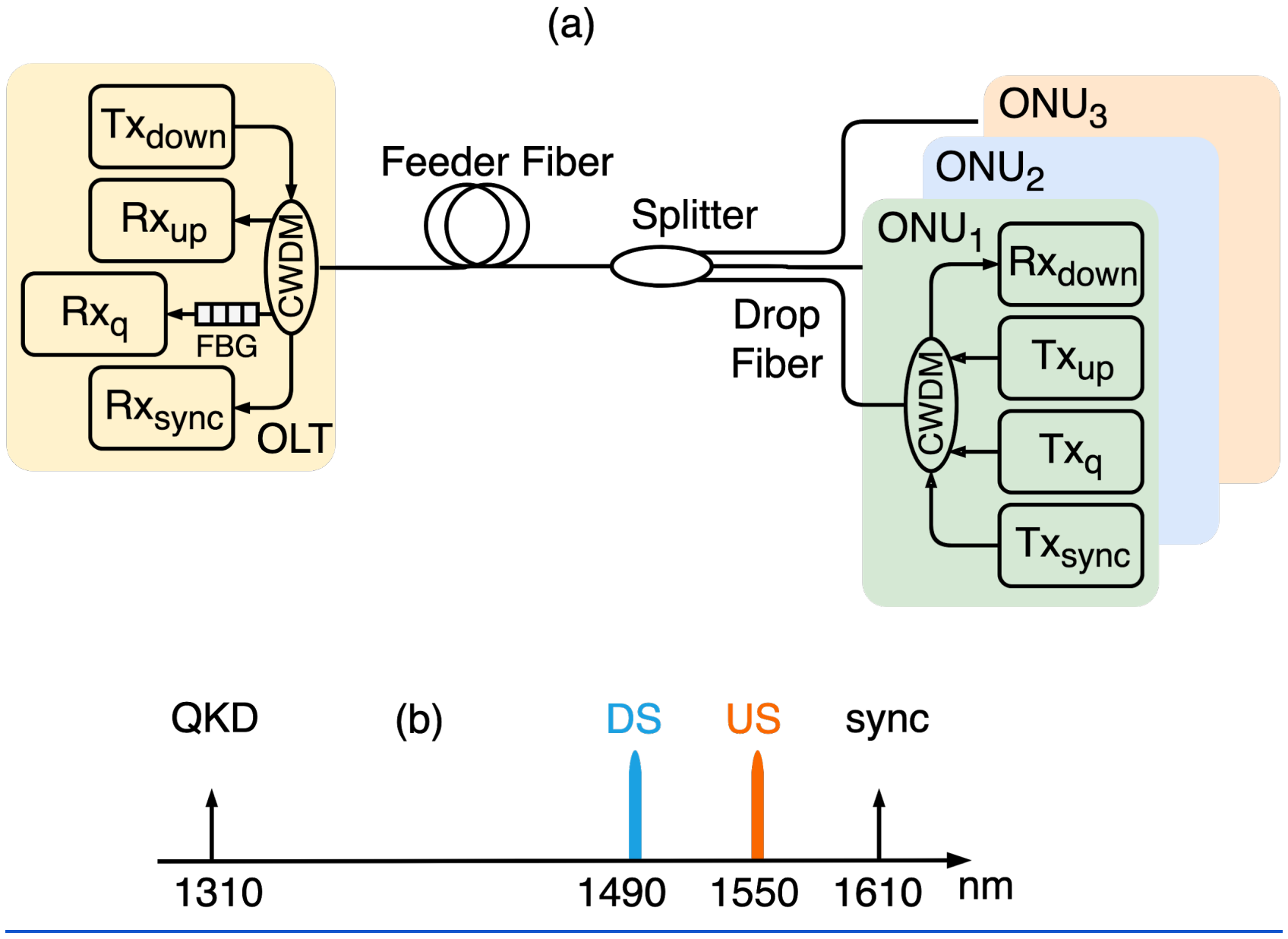

Fig.3. (a) TDM-PON with integrated quantum and SYNC channels. (b) Wavelength plan. Classical DS, US, quantum, and SYNC channels use four different wavelengths.

Although the quantum channel is far away from classical channels, Raman scattering noise is still the dominant factor limiting the key rate and fiber distance of QKD. Since quantum channels co-propagates with US channels, Raman noise originates from backscattering of the DS channel and forward scattering of US channels. Since Raman scattering is proportional to the incident optical power and backscattering has higher efficiency than forward scattering, backscattering of DS channel in feeder fiber is the dominating noise source. This is because: 1. DS channel in the feeder fiber has the highest optical power; 2. backscattering noise generated in the feeder fiber can reach the quantum Rx in OLT without going through the attenuation of splitter; 3. in drop fibers, DS optical power is attenuated by the splitter; 4. backscattering noise generated in drop fibers is attenuated by the splitter one more time before reaching the quantum Rx. The contribution of forward scattering of US channels is smaller because of the lower power of US channels. In drop fibers, each ONU only has limited time slot to transmit US data; in feeder fiber, the US optical power is attenuated by the splitter.

In terms of user number scalability, a power splitter with higher split ratio is needed to support more users, which introduces higher loss to the quantum channel. However, the dominant backscattering noise generated in the feeder fiber is unattenuated since it does not have to pass through the splitter to reach the quantum Rx. For a large number of users, it is challenging to integrate QKD into TDM-PON using the architecture in Fig. 3.

In terms of distance scalability, a more detailed investigation reveals that backscattering noise increases with fiber length until saturation; whereas forward scattering noise first increases then decreases with fiber length and has a peak at ~20 km. In TDM-PON, QKD over 20 km distance, including both feeder and dropper fibers, is only possible if extra mitigation techniques are exploited to mitigate Raman noise.

There are three techniques to mitigate the effect of Raman noise, i.e., wavelength filtering, temporal filtering, and power control of classical channel. Besides the CWDM in Fig. 3(a), an additional narrow filter based on fiber Bragg grating (FBG) or even multiple stages of filters can be added before the quantum Rx to eliminate out-of-band Raman noise. Temporal filtering makes SPDs work in gate mode, so that only photons within the detection window will be received. In real PON operation, QKD is used to exchange master keys between OLT and ONUs. The quantum channel for each ONU only needs to be turned on for a short period of time for key updating. Depending on the security level, updating frequency of master key is quite low, once every few hours or days. Due to the low frequency and short operation period, it is feasible to reduce the optical power of classical channels or even turn them off during the key updating period.

Fig. 4 shows a TDM-PON architecture, where classical channels are turned off during QKD operation to avoid Raman noise. The optical switches in the OLT and all ONUs are synchronized to switch the system between quantum and classical modes. In the quantum mode, classical DS and US channels are turned off, so quantum and SYNC channels can reuse their wavelengths. Like the upstream of TDM-PON, quantum channels from all ONUs are time interleaved. At any given time, the quantum Rx in OLT only receives photons from one ONU. In this system, the Raman noise is only contributed by the forward scattering of SYNC channels. For each user, only its own SYNC channel contributes to the Raman noise of this user's quantum channel. Since SYNC channels have much lower power and data rate than DS and US channels, Raman noise in this system is much less than Fig. 3.

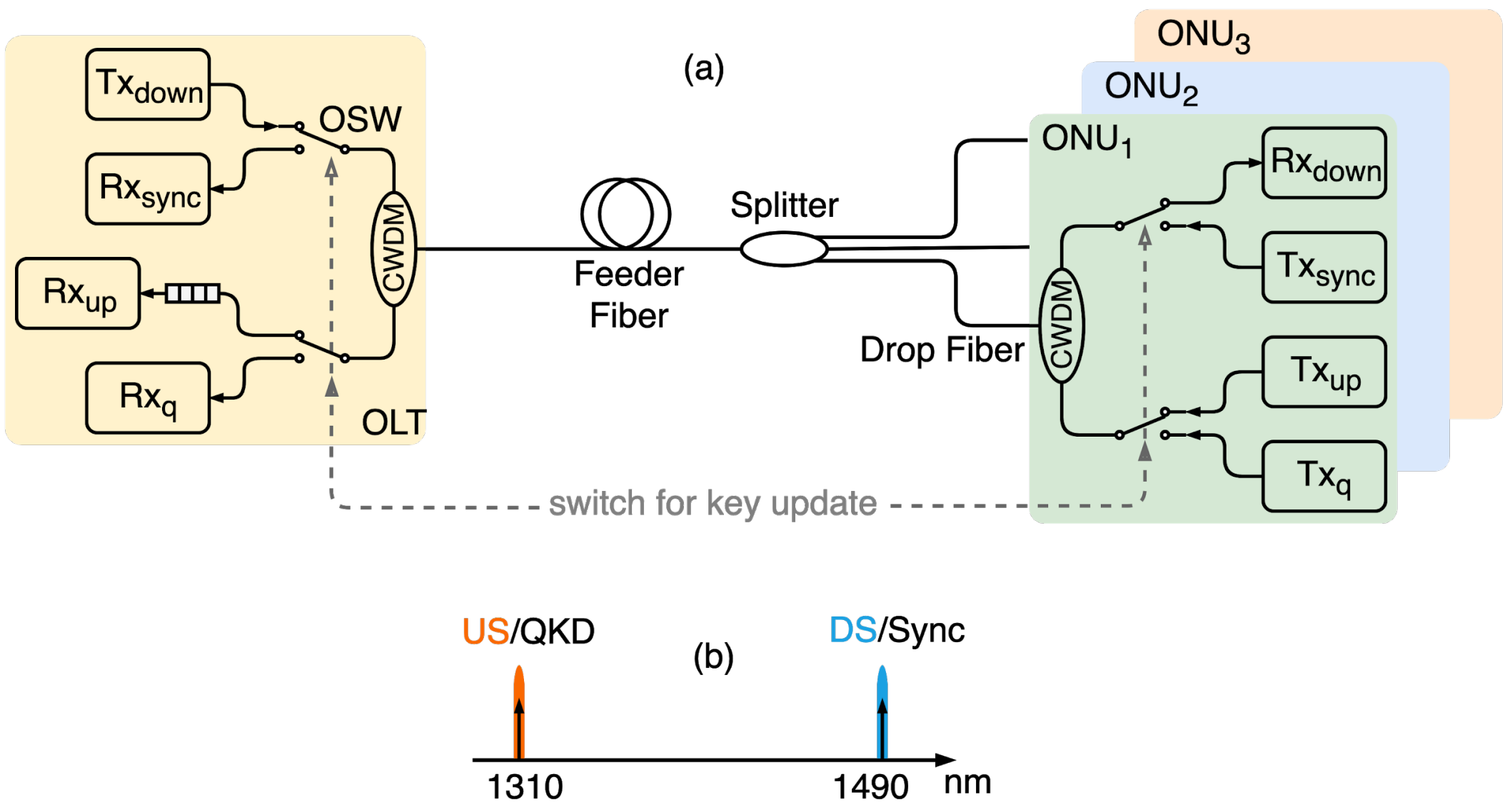

Fig. 4. (a) TDM-PON with integrated quantum and SYNC channels. Optical switches make the system work in quantum and classical modes alternatively. (b) Wavelength plan. Quantum and SYNC channels reuse the wavelengths of US and DS channels.

**Integration of QKD with WDM-PON**

In WDM-PON, each ONU has dedicated wavelengths. Compared with TDM-PON, the absence of the power splitter reduces the attenuation of quantum channels and increases the QKD distance. Fig. 5 shows the architecture and wavelength plan of integrating QKD into a WDM-PON. Each ONU uses four wavelengths in O, S, C, and L bands for quantum, classical DS, US, and SYNC channels respectively. They are multiplexed by a CWDM and fed to a drop fiber. All wavelengths coming from the drop fibers of all ONUs are bundled together by a cyclic arrayed waveguide grating (AWG), then transmitted via a feeder fiber to the OLT. The cyclic AWG can perform wavelength multiplexing in O, S, C, and L bands simultaneously. There are 4N wavelengths in the feeder fiber, within which 2N channels are high-power classical channels. N is the number of users. At OLT, a CWDM first separates the four groups of wavelengths. For each group, an AWG further separates the channels for each user.

One drawback of this architecture is the cyclic AWG to multiplex wavelengths in four bands simultaneously, which is not commercially available. Moreover, ITU-T DWDM grids are only defined in C and L bands, but not available in O and S bands, which also increases the cost of AWGs used in OLT.

In TDM-PON, the dominant noise to quantum channels is the Raman backscattering of DS channels in feeder fiber. This is also true for WDM-PON. Raman noise generated in drop fibers is negligible compared with that originated in feeder fiber. This is because in the feeder fiber the scattering noise is contributed by all DS and US channels; while in drop fibers, it is only contributed by one DS and one US channels. Second, due to the filtering effect of cyclic AWG, in the drop fiber of i-th user, only those Raman noise with the same wavelength of this user's quantum channel can pass through the AWG and reach its quantum receiver. All other noise in the drop fiber at different wavelengths is blocked by the cyclic AWG. But in feeder fiber, any noise located in the band of the quantum channels can pass through the CWDM and AWG in OLT, and reach some users' quantum receivers. Compared with forward scattering of US channels, DS channels have higher optical power, especially in feeder fiber. Therefore, for the WDM-PON in Fig. 5, dominant noise source is the backscattering of DS channels in feeder fiber.

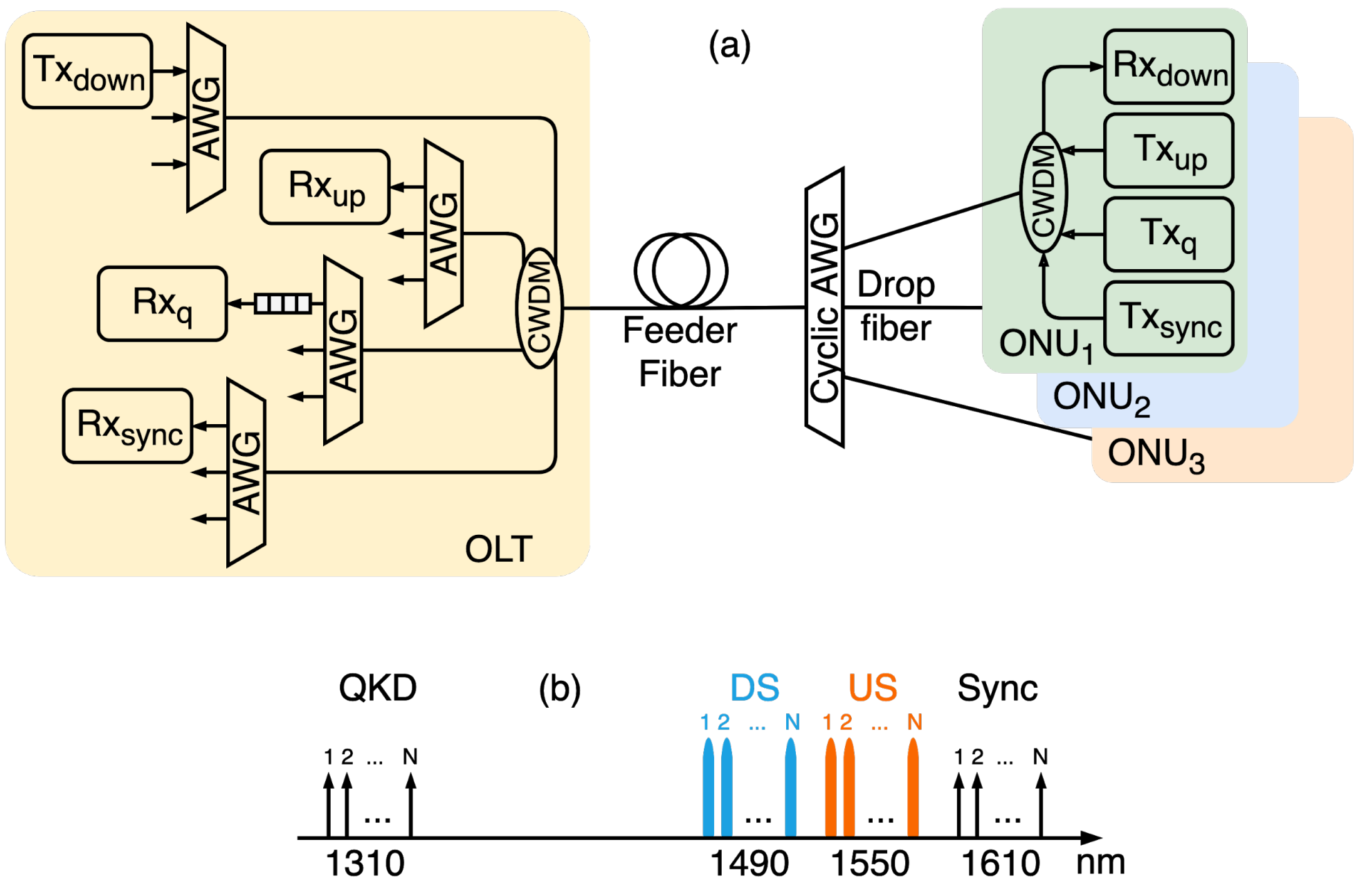

Fig. 5. (a) WDM-PON with integrated quantum and SYNC channels. Each ONU uses four wavelengths for quantum, classical DS, US, and SYNC channels. (b) Wavelength plan. In feeder fiber, there are 4N wavelengths and 2N classical DS/US channels. N is the user number.

Since there are 2N classical channels in the feeder fiber, significant Raman scattering is expected, and quantum channels may not work without extra techniques to reduce Raman noise. Besides wavelength filtering, temporal filtering, and power control of classical channels, two variants of the system in Fig.5 is proposed, as shown in Fig. 6 and 7.

Fig. 6 has the same wavelength plan as Fig. 5 but uses two feeder fibers. Instead of separating four groups of wavelengths in OLT, the quantum channels (O-band) of all users are taken out first by a CWDM after the cyclic AWG and transmitted via a dedicated feeder fiber. All classical channels (DS, US, and SYNC) are carried by another feeder fiber. Using the dual-feeder fiber architecture, the dominant noise source in the system of Fig. 5, i.e., backscattering of DS channels in feeder fiber, is eliminated. The residual Raman noise only comes from drop fibers. Moreover, thanks to the filtering effect of cyclic AWG, the Raman noise to one user's quantum channel is only generated by its own DS and US channels in its own drop fiber. Noise from other drop fibers is all blocked by the cyclic AWG. Therefore, the dual-feeder fiber architecture in Fig. 6 significantly reduces Raman noise.

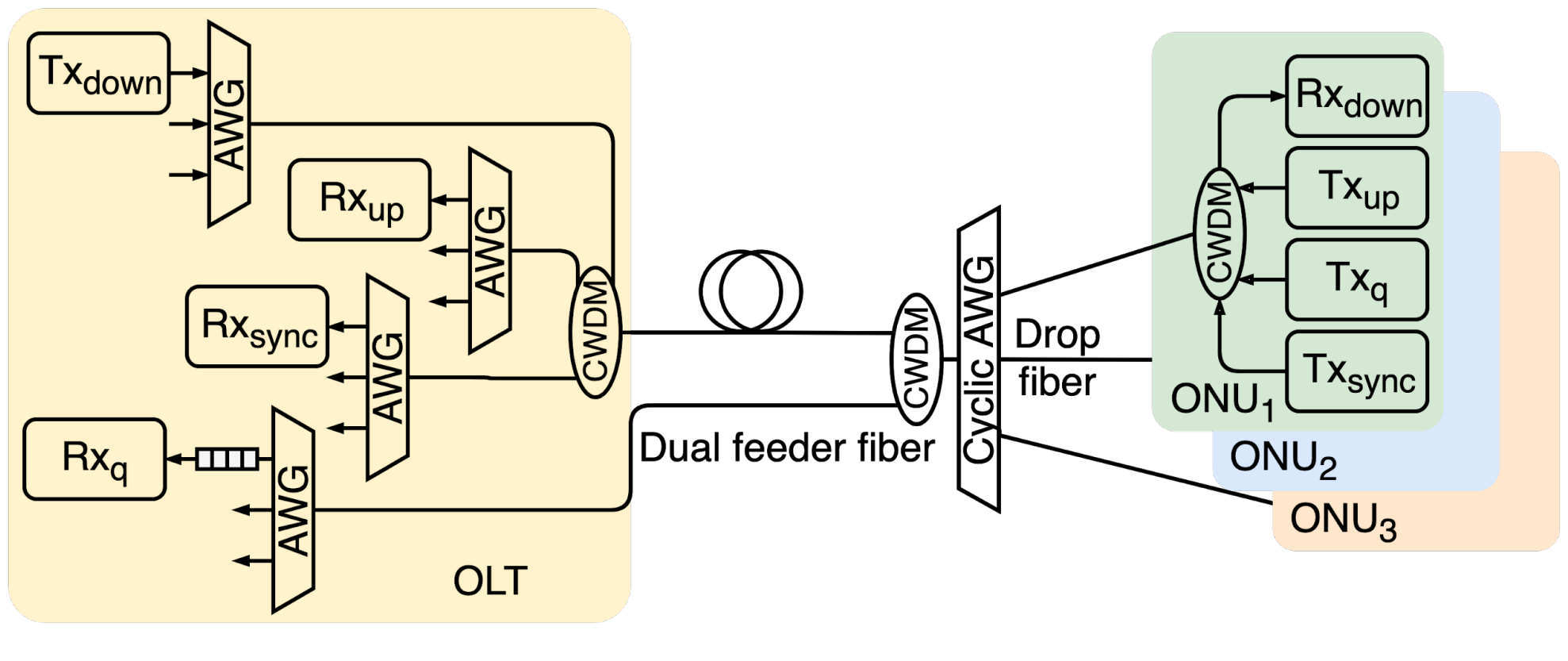

Fig. 6. WDM-PON with integrated QKD using a dual-feeder fiber architecture.

The single-feeder fiber system also has poor scalability for large number of users, since the Raman noise increases with user number. For dual-feeder fiber system, however, since the noise source to each user's quantum channel is only contributed by its own DS and US channels in its own drop fiber, adding more users will not increase the Raman noise, which significantly improves the network scalability. Adding a second feeder fiber for quantum channels increases the system cost. However, the cost increment is shared by multiple users.

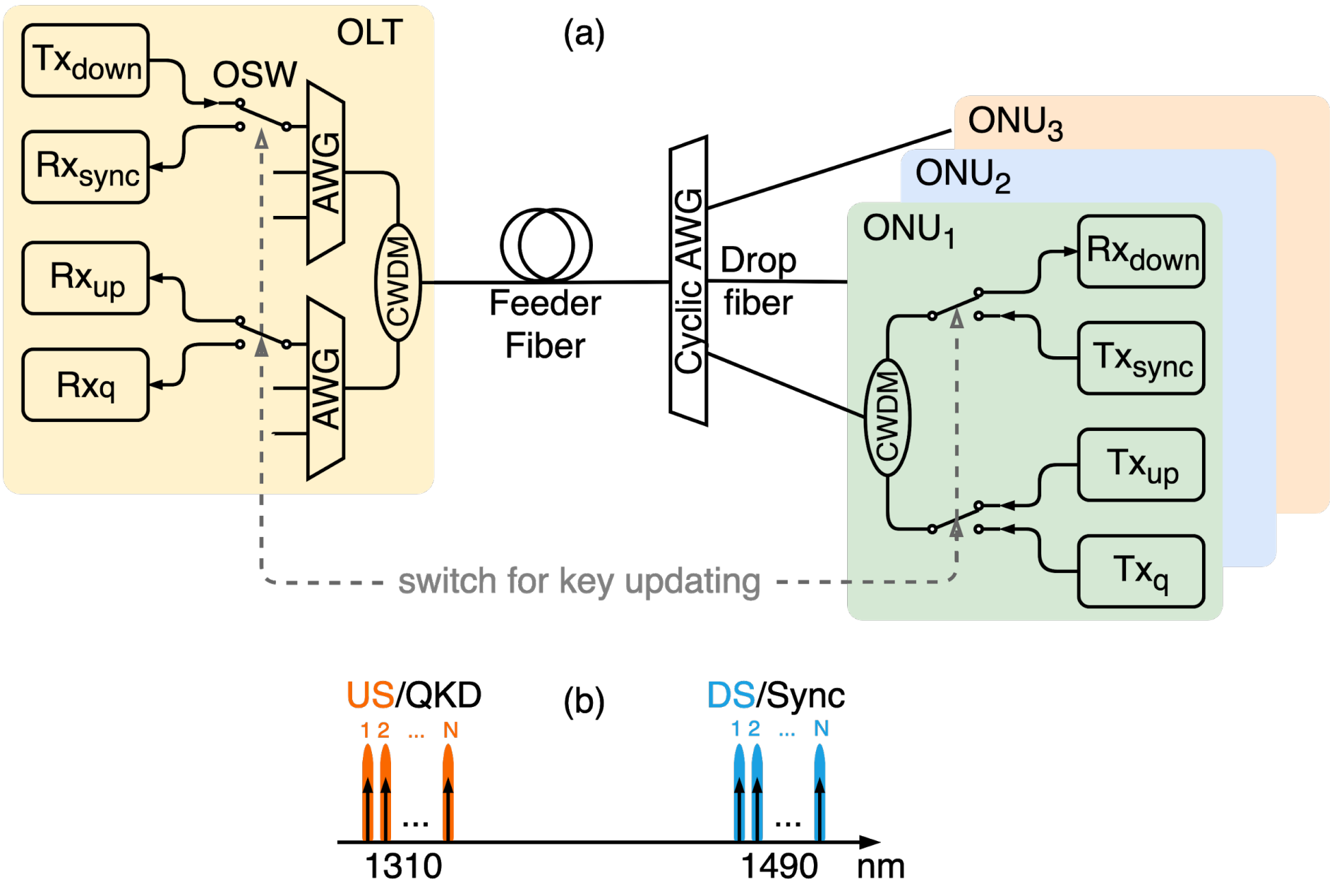

Fig. 7. (a) WDM-PON with integrated quantum and SYNC channels. Optical switches make the system work in quantum and classical modes alternatively. (b) Wavelength plan. Quantum and SYNC channels reuse the wavelengths of US and DS channels.

Fig. 7 shows a WDM-PON architecture, where classical channels are turned off during QKD operation to avoid Raman noise. The optical switches in the OLT and all ONUs are synchronized to switch the system alternatively between quantum and classical modes. In the quantum mode, classical DS and US channels are turned off, quantum and SYNC channels reuse their wavelengths. The only Raman noise is contributed by SYNC channels. Since the optical power of SYNC channels are much lower than DS and US channels, Raman noise in this system is much lower than that in Fig. 5.

## Discussion

A comparison of the pros and cons of the systems in Fig. 3-7 is shown in Table 1. It should be noted that all designs proposed in this paper are agnostic to QKD protocols, and can work with various protocols, such as polarization-coded or phase-coded decoy-state BB84, T12, coherent one way, without modification of the network architecture.

In terms of compatibility with quantum channels, classical channels using coherent optics are preferred to those using intensity modulation/direct detection (IM/DD). This is because spontaneous Raman scattering noise is proportional to optical launch power and the bandwidth of the incident light. To convey the same amount of data, a coherent channel occupies less bandwidth than IM/DD channels due to its higher spectral efficiency. Moreover, thanks to the intradyne detection with an optical local oscillator, a coherent channel has higher receiver sensitivity and needs less optical launch power than IM/DD channels.

Table 1. Comparison of different TDM/WDM-PON architectures

| PON | TDM | | WDM | | |
|---|---|---|---|---|---|
| Figure | 3 | 4 | 5 | 6 | 7 |
| Wavelength number | 4 | 2 | 4N | 4N | 2N |
| Noise source | Backscattering of DS channel in feeder fiber | Forward scattering of SYNC channels | Backscattering of DS channels in feeder fiber | Raman scattering in drop fibers | Forward scattering of SYNC channels |
| Noise mitigation techniques | No | Mode switching between quantum/classical | No | Dual feeder fiber | Mode switching between quantum/classical |
| Pros | | Reduced Raman noise | | Reduced Raman noise<br>Raman noise independent on user number | Reduced Raman noise |
| Cons | High Raman noise<br>Poor scalability for many users | Switching between two modes | Four-band cyclic AWG<br>Large wavelength number<br>High Raman noise<br>Poor scalability for many users | Four-band cyclic AWG | Switching between two modes |

* N is the number of users

## Conclusion

In this paper, we proposed systems and methods to integrate QKD into existing TDM and WDM-PONs to enable the key delivery to end users for end-to-end encryption. To address the P2MP topology and fiber deficiency of PON, we proposed network architectures and wavelength plans to multiplex quantum channels with classical channels without the necessity to deploy new fibers. The main challenge to the coexistence of quantum and classical channels is the interference due to the Raman scattering noise generated by classical channels. For TDM and WDM-PONs, we thus proposed an alternative operation mode between the classical and quantum channels to alleviate the Raman noise. For WDM-PON, a dual-feeder fiber architecture is also proposed to mitigate Raman noise.

## Acronym

| | |
|---|---|
| AES | advanced encryption standard |
| AWG | arrayed waveguide grating |
| CWDM | coarse wavelength division multiplexer |
| DS | downstream |
| FBG | fiber Bragg grating |
| IM/DD | intensity modulation/direct detection |
| OLT | optical line terminal |
| ONU | optical network unit |
| P2MP | point-to-multipoint |
| P2P | point-to-point |
| PKI | public key infrastructure |
| PON | passive optical network |
| QKD | quantum key distribution |
| Rx | receiver |
| SPD | single photon detector |
| SYNC | synchronization |
| TDM | time division multiplexing |
| TDMA | time division multiple access |
| Tx | transmitter |
| US | upstream |
| WDM | wavelength division multiplexing |
| XOR | exclusive or |